\documentclass[12pt,aps]{revtex4}
\usepackage{epsf}

\def\beq{\begin{equation}}
\def\eeq{\end{equation}}
\def\beqn{\begin{eqnarray}}
\def\eeqn{\end{eqnarray}}
\def\bea{\begin{eqnarray}}
\def\eea{\end{eqnarray}}
\def\be{\begin{equation}}
\def\ee{\end{equation}}

\begin{document}

\voffset 1.25cm

\title{$W^- \rightarrow \tau \bar \nu_\tau$ 3-sigma anomaly
in new physics beyond the standard model}
\author{Shou-hua Zhu}
\affiliation{Institute of Theoretical Physics, School of Physics \\
Peking University, Beijing 100871, China}

\date{\today}

\begin{abstract}
Among so-called three 3-sigma anomalies in high energy
physics, the excess of the
branching ratio $W^- \rightarrow \tau \bar \nu_\tau$ with respect
to the electrons and muons is especially interesting because (1)
in the standard model (SM), $W^-\ell\bar \nu_\ell$ is the pure left-handed
charge-current which has been tested precisely already, at least
for the first two generation fermions, and (2) the $W^\pm$
two-body leptonic decay is the cleanest one among three anomalies
due to its simpler kinematics and less hadronic uncertainties. In
this paper, we explore the possibilities to account for the
anomaly in type II two-Higgs-doublet model (2HDM) and minimal
supersymmetric model (MSSM), as well as effective lagrangian
approach by introducing anomalous left- and right-handed $W^-\tau
\bar\nu_\tau$ couplings. Our results show that 2HDM and MSSM can
hardly accommodate $W^- \rightarrow \tau \nu_\tau$ anomaly, and
the anomaly is only marginally consistent to the measurements of
$\tau \rightarrow \nu_\tau \ell \bar \nu_\ell$ at 95\% confidence
level with the presence of anomalous couplings.
In the allowed regions, the right-handed coupling of
$W^-\tau \nu_\tau$ shifts from $0$ in SM
to $\sim 0.12$ while the left-handed one from 1 to $\sim 1.005$.

\end{abstract}


\maketitle

\newpage

The standard model (SM) of high energy physics can successfully
describe all experiments at LEP, SLD and Tevatron etc. at one-loop
level, except three so-called 3-sigma anomalies: spread in
$\sin^2\theta_{eff}$ at the Z pole, NuTeV and W branching
fractions \cite{LEP-latest-talk}. The last one is especially
interesting because (1) in the SM, $W^-\ell\bar \nu_\ell$ is the
pure left-handed charge-current which has been tested precisely
already, at least for the first two generation fermions, and (2)
the $W^\pm$ two-body leptonic decay is the cleanest one among
three anomalies due to its simpler kinematics and less hadronic
uncertainties. In the SM, the mass effects of leptons in $W^\pm$
leptonic two-body decays are negligible at LEP energy. Therefore,
the branching fractions for electron, muon and tau decays should
be the same. However the measurements from LEP \cite{:2004qh} show
that an excess of the branching ratio $W^- \rightarrow \tau \bar
\nu_\tau$ with respect to the other leptons is evident. The excess
can be quantified with the two-by-two comparison of these
branching fractions as
\begin{eqnarray}
\frac{Br( W^- \rightarrow \mu
\bar \nu_\mu)}{Br( W^- \rightarrow e^- \bar \nu_e) }
&=& 0.994 \pm 0.020 \label{ue} \\
\frac{Br( W^- \rightarrow \tau \bar \nu_\tau)}{Br( W^- \rightarrow
e^- \bar \nu_e)} &=& 1.070 \pm 0.029 \label{taue}\\
\frac{Br( W^- \rightarrow \tau \bar \nu_\tau)}{Br( W^- \rightarrow
\mu \bar \nu_\mu)} &=& 1.076 \pm 0.028.\label{taumu}
\end{eqnarray}
While the branching fractions of W into electrons and muons
perfectly agree, the branching fractions in taus with respect to
electrons and muons differ
 by more
than two standard deviations, where correlations have been taken into
account.
The ratio between the tau fractions and the average of electrons and muons can
be computed:
\begin{eqnarray}
\frac{2 Br( W^- \rightarrow \tau \bar \nu_\tau)}{Br( W^-
\rightarrow \mu \bar \nu_\mu)+Br( W^- \rightarrow e^- \bar \nu_e)}
=1.073 \pm 0.026. \label{taumue}
\end{eqnarray}

Before we proceed further, it is worthwhile to mention that the
precise tests of neutral weak current at LEP experiments at Z-pole
have reached to an accuracy of $O(0.1\%)$. Therefore it is
challenging to find a solution to account for both $W \rightarrow
\tau \bar\nu_\tau$ anomaly and $Z \rightarrow \tau \bar \tau$.
Obviously new physics of oblique-type, i.e. new physics
contributions enter only via vacuum polarization effects to gauge
boson propagators of the SM, can hardly explain $W^- \rightarrow
\tau \bar \nu_\tau$ anomaly because the corrections are the same
for $\tau \bar \nu_\tau$ and $\mu \bar \nu_\mu$. The natural
source for $W^- \rightarrow \tau \bar \nu_\tau$ anomaly is the
flavor-dependent yukawa interactions among Higgs and fermions in
multi-Higgs models such as type II two-Higgs-doublet model (2HDM)
and minimal supersymmetric standard model (MSSM). In MSSM, the
flavor-dependent interactions also exist among chargino
(neutrolino) slepton and lepton.  Especially the well-known
$\tan\beta$ enhancement of yukawa couplings of the third family
fermions may play a special role to accommodate W branching
fractions anomaly. In this paper, we will firstly explore whether
such kind of flavor-dependent interactions can account for the
$W^- \rightarrow \tau \bar \nu_\tau$ anomaly in the popular type
II 2HDM and MSSM. Our numerical results show that these two models
can hardly explain anomaly. Therefore we then study this issue
under the framework of effective lagrangian approach. By introducing
anomalous $W\tau \bar\nu_\tau$ left- and right-handed couplings,
the anomaly can be accommodated. However, the allowed parameters
are only marginally consistent to the limits from measurements
from
$\tau \rightarrow \nu_\tau \ell \bar \nu_\ell$.\\

In order to gauge the new physics contributions, we define
\begin{eqnarray}
\delta_{new}\equiv \frac{\Gamma^{NLO}-\Gamma^{NLO,SM}}{\Gamma^0}
\end{eqnarray}
where $\Gamma^{NLO}$, $\Gamma^{NLO,SM}$, $\Gamma^0$ are the decay widths
at NLO in new models,  at NLO in the SM, at tree-level in the SM
respectively.
Assuming the flavor-dependent interactions have negligible effects
on $W^- \rightarrow \mu \bar\nu_\mu$ and
$W^- \rightarrow e \bar\nu_e$, from eq. \ref{taumue}, we obtain
\begin{eqnarray}
\delta_{new}= 0.073 \pm 0.026.
\label{limit}
\end{eqnarray}
In the following we will study whether the new physics is
compatible with such large effects.

In Ref.\cite{Lebedev:2000ix}, the authors studied
lepton universality violation in W decay in general 2HDM for large
$\tan\beta$. From their results,
the two-Higgs doublet model usually predicts a decrease of Br($W^-
\rightarrow \tau \bar \nu_\tau$), except in the limit the Higgs
mass splitting is small ($\le m_W/2$). Setting
$m_{A^0}=m_{H^0}=m_{H^\pm}\equiv m$ and neglecting $m_{h^0}$
contributions, we can express
\begin{eqnarray}
\delta_{new}=\left(\frac{g m_\tau \tan\beta}{6 \pi
m}\right)^2.
\end{eqnarray}
For $\tan\beta=100$ and $m=200$ GeV
\begin{eqnarray}
\delta_{new} \approx 0.001.
\end{eqnarray}
Obviously the general 2HDM can hardly explain $W^- \rightarrow
\tau \bar \nu_\tau$ anomaly.

In MSSM, besides the enhanced Higgs-fermions couplings, the
chargino and neutrolino interactions with lepton and slepton
contain also $m_\tau \tan\beta$ terms. In our analytical and
numerical calculations, we use Feynarts, FormCalc and LoopTools
\cite{Hahn:2000jm} to evaluate the mass, mixing angle etc. from
input parameters, at the same time enforce the experimental
constraints, for examples the chargino, neutrolino and stau mass
lower limits as well as $\rho$ parameter. Moreover we use these
packages to calculate the decay width for both $W^- \rightarrow
\tau \bar \nu_\tau$ and $Z \rightarrow \tau \bar \tau$ in the
MSSM.

We have scanned the whole parameter space, and find that in the allowed
parameter regions
\begin{eqnarray}
\delta_{new} \le 0.001.
\end{eqnarray}
It should be noted that for the large region of parameter space,
$\delta_{new}$ is negative, which is the same with that in 2HDM.
In the following we show the results of positive $\delta_{new}$
for certain typical parameters. In Fig. 1-3, we depict
 $\delta_{new}$ as functions of $\tan\beta$, $\mu$ for
$W^- \rightarrow \tau \bar \nu_\tau$ and as a function of
$\tan\beta$ for $Z \rightarrow \tau \bar \tau$. For these three
figures, $m_{\tilde{\tau}_1}=90$ GeV, $\mu=100$ GeV,
$m_{A^0}=M_2=1$ TeV, and other sfermion masses are taken as 1 TeV.
From figures we can see clearly that in the allowed parameters in
MSSM, it is impossible to account for $W^- \rightarrow \tau \bar
\nu_\tau$ anomaly. Moreover the comparable $\delta_{new}$ in $W^-
\rightarrow \tau \bar \nu_\tau$ and $Z \rightarrow \tau \bar \tau$
indicate that it is very hard to account for both charged- and
neutral-currents data simultaneously.

Now we switch to the effective lagrangian approach.
In this paper, we
simply explore the anomalous couplings only for $ W\tau \bar \nu_\tau$ sector
which are written as
\begin{eqnarray}
L= \frac{g}{\sqrt{2}} W^\mu \bar\nu \gamma_\mu \left( (1+\delta_L) P_L
+\delta_R P_R \right) \tau+ h.c.
\end{eqnarray}
where $P_{L,R}=1/2(1\mp \gamma_5)$ and $\delta_L=\delta_R=0$ for the
SM case.

It is straightforward to write the constraint based on eq. \ref{taumue} as
\begin{eqnarray}
(1+\delta_L)^2+\delta_R^2=1.073\pm 0.026.
\label{lrc1}
\end{eqnarray}
Limits of anomalous couplings $\delta_L$ and $ \delta_R$ can also
be derived from precise measured Michel parameters which are
extracted from the energy spectrum of the charged daughter lepton
$\ell=e, \mu$ in the decays $\tau \rightarrow \nu_\tau \ell \bar
\nu_\ell$ \cite{Eidelman:2004wy}. We can write the limits as
\cite{Eidelman:2004wy}
\begin{eqnarray}
|1+\delta_L| < 1.005 \ {\rm and} \ |\delta_R| <0.12 \ \ \ {\rm at} \ 95\% \
{\rm CL}.
\label{lrc2}
\end{eqnarray}
The allowed regions at 95\%
confidence level from eqs. \ref{lrc1} and \ref{lrc2} are shown in Fig.4.
From the figure we can see that
the allowed regions are severely constrained and
measurements of $W^- \rightarrow \ell \bar \nu_\ell$ only marginally
agree to those of  $\tau \rightarrow \nu_\tau \ell \bar \nu_\ell$.
In the allowed regions, the right-handed coupling shifts from $0$ in SM
to $\sim 0.12$ while the left-handed one from 1 to $\sim 1.005$.\\

To summarize, in the popular type II 2HDM and MSSM, the
$\tan\beta$ enhanced flavor-dependent yukawa interactions have
little impact on $W^- \rightarrow \tau \nu_\tau$ anomaly. If the
anomaly stands up to the scrutiny of the future high energy
physics at LHC and ILC, there must be new physics other than
general 2HDM and MSSM, for examples gauge models of generation
non-universality \cite{Li:1981nk,LiPrivate}. By introducing
anomalous left- and right-handed couplings of
$W^-\tau\bar\nu_\tau$, we explore the anomaly under effective
lagrangian approach. Our results show that  $W^- \rightarrow \tau \nu_\tau$
anomaly is only marginally consistent to the measurements of
$\tau \rightarrow \nu_\tau \ell \bar \nu_\ell$ at 95\% confidence level.
In the allowed regions, the right-handed coupling shifts from $0$ in SM
to $\sim 0.12$ while the left-handed one from 1 to $\sim 1.005$.\\


\noindent {\em Acknowledgements}: The author thanks X.Y. Li and
H.Q. Zheng for the stimulating discussions and X.Y. draw the $W\rightarrow
\tau\bar \nu_\tau$ anomaly into my attention. This work was
supported in part by the Natural Sciences Foundation of China
under grant no. 90403004, and computing facility in Carleton University.

\begin{figure}
\epsfxsize=15 cm \centerline{\epsffile{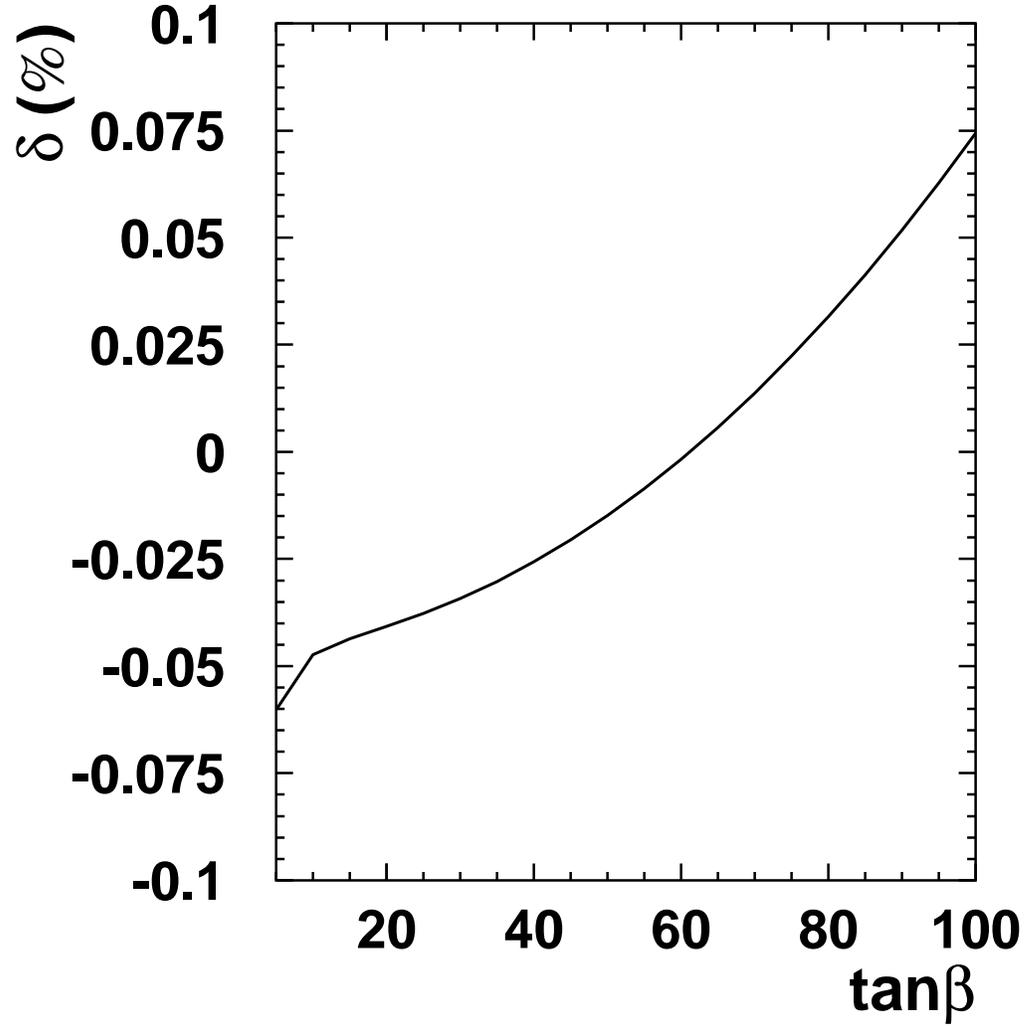}}
\caption{$\delta_{new}$
for $W^- \rightarrow \tau \bar \nu_\tau$ as a function of $\tan\beta$ where
$m_{\tilde{\tau}_1}=90$ GeV, $\mu=100$ GeV, $m_{A^0}=M_2=1$ TeV,
and other sfermion masses are taken as 1 TeV. }
\label{fig1}
\end{figure}

\begin{figure}
\epsfxsize=15 cm \centerline{\epsffile{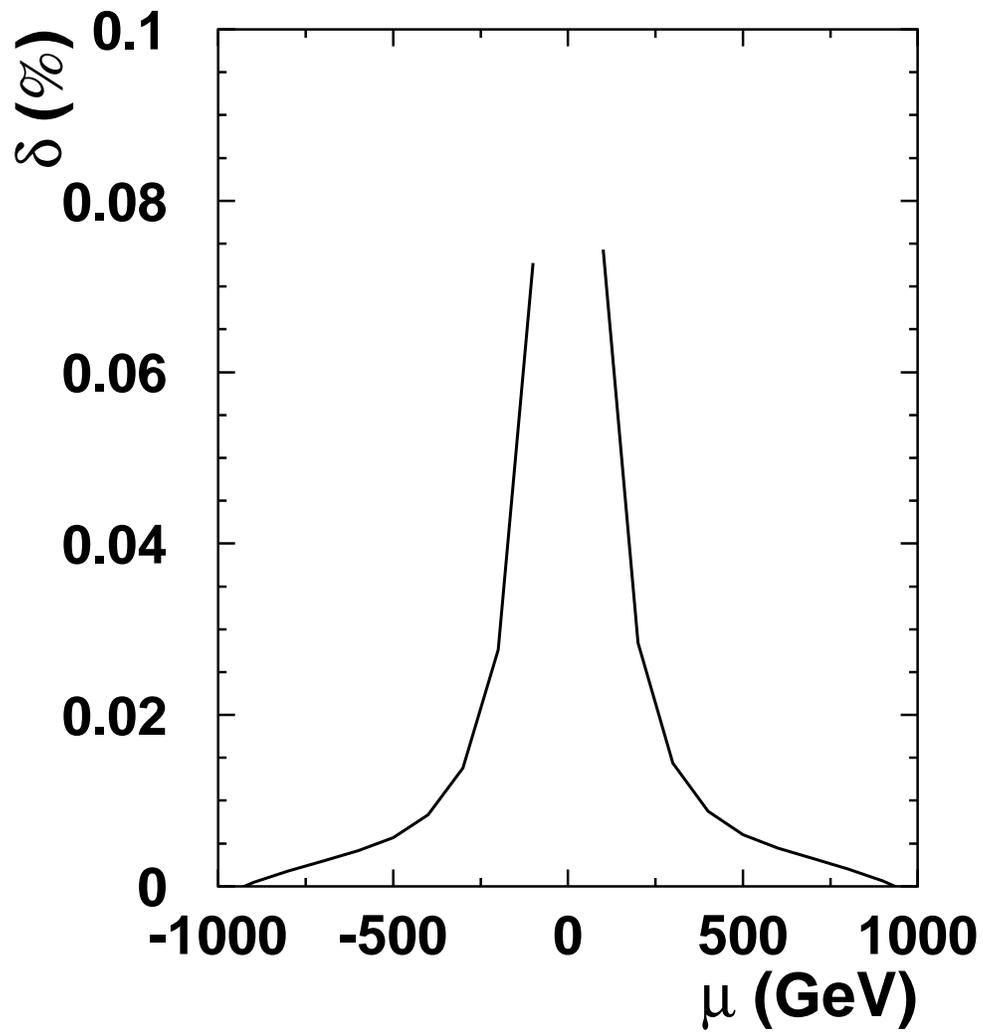}}
\caption{$\delta_{new}$
for $W^- \rightarrow \tau \bar \nu_\tau$ as a function of $\mu$
with $\tan\beta=100$. Other parameters are the same with Fig. \ref{fig1}.  }
\label{fig2}
\end{figure}

\begin{figure}
\epsfxsize=15 cm \centerline{\epsffile{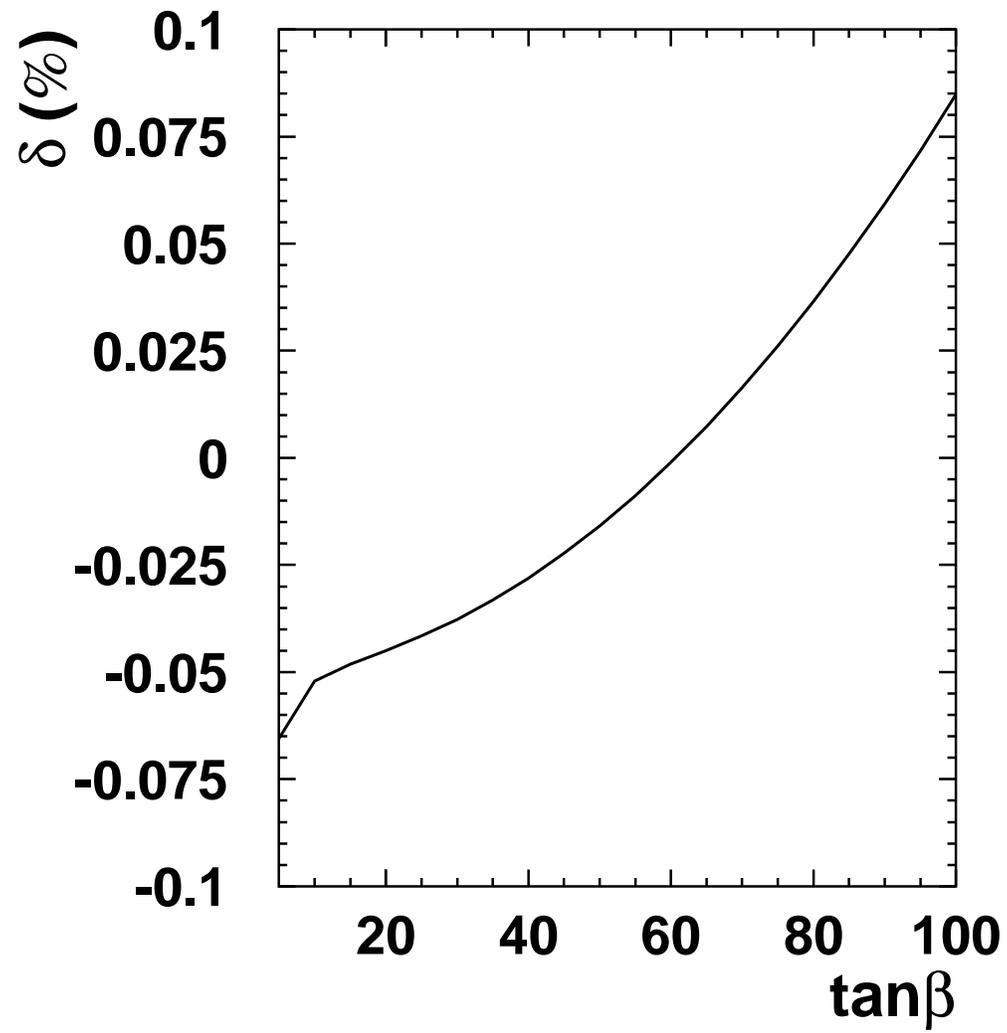}}
\caption{$\delta_{new}$
for $Z \rightarrow \tau \bar \tau$ as a function of $\tan\beta$.
Other parameters are the same with Fig. \ref{fig1}.}
\label{fig3}
\end{figure}

\begin{figure}
\epsfxsize=15 cm \centerline{\epsffile{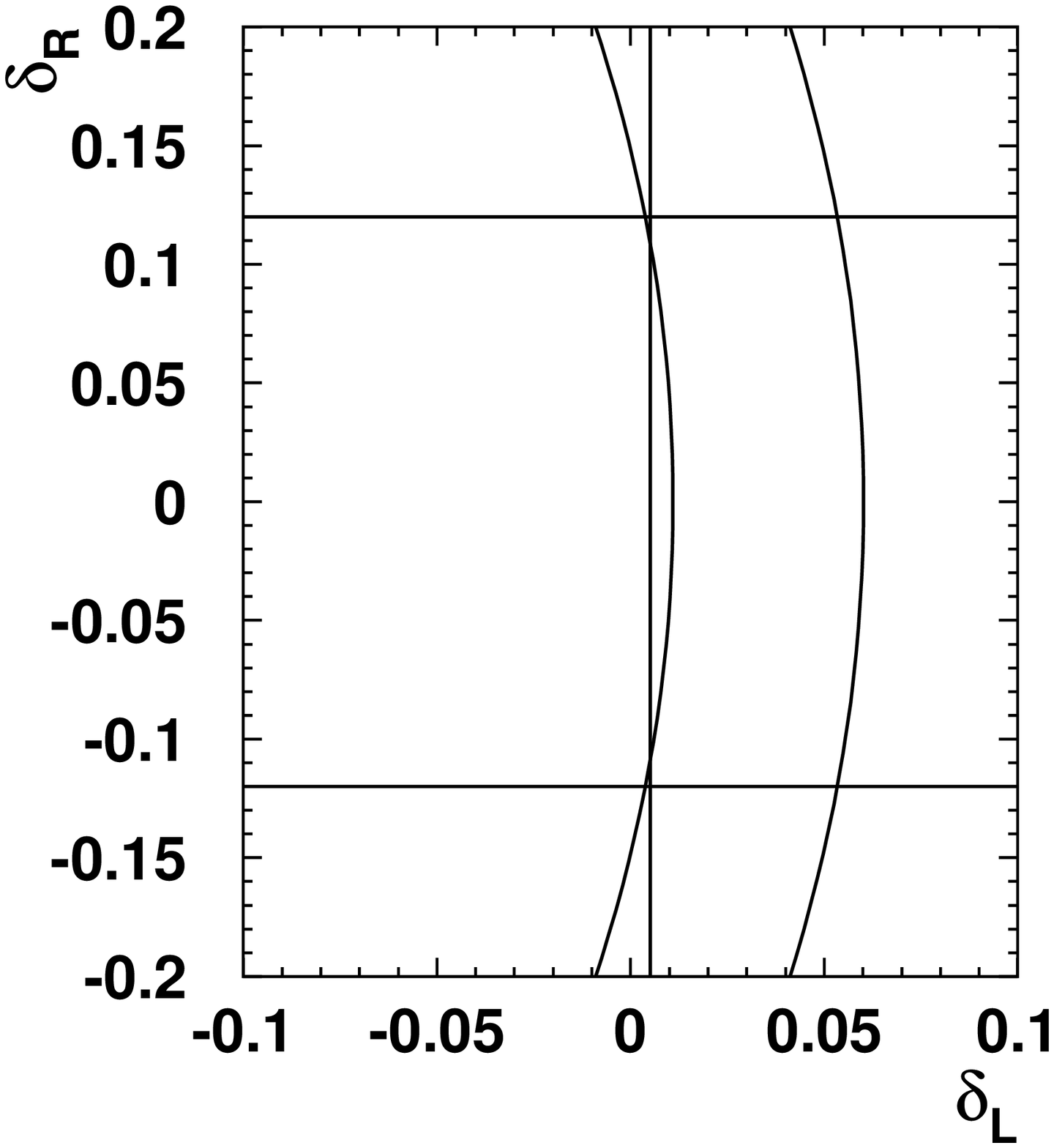}} \caption{
Allowed regions of $\delta_L$ and $\delta_R$ at 95\% confidence
level constrained by the measurements of $\tau \rightarrow
\nu_\tau \ell \bar \nu_\ell$ (between two parallel lines and to
the left of vertical line) as well as $W^- \rightarrow \ell \bar
\nu_\ell$ (between two arcs). } \label{fig4}
\end{figure}


\begin{thebibliography}{99}

\bibitem{LEP-latest-talk}
For latest talk see, Martin Gruenewald at IOP HEPP 2005, March
2005. {\rm
http://lepewwg.web.cern.ch/LEPEWWG/misc/mwg\_iop05.pdf}.

\bibitem{:2004qh}
    [LEP Collaborations],
  arXiv:hep-ex/0412015.

\bibitem{Lebedev:2000ix}
  O.~Lebedev, W.~Loinaz and T.~Takeuchi,
  Phys.\ Rev.\ D {\bf 62}, 055014 (2000)
  [arXiv:hep-ph/0002106] and references therein.

\bibitem{Hahn:2000jm}
  T.~Hahn,
  Nucl.\ Phys.\ Proc.\ Suppl.\  {\bf 89}, 231 (2000)
  [arXiv:hep-ph/0005029].


\bibitem{Li:1981nk}
  X.Y.~Li and E.~Ma,
  Phys.\ Rev.\ Lett.\  {\bf 47}, 1788 (1981).

\bibitem{LiPrivate}

X.Y. Li, private communication.

\bibitem{Eidelman:2004wy}
  S.~Eidelman {\it et al.}  [Particle Data Group],
  Phys.\ Lett.\ B {\bf 592}, 1 (2004).

\end{thebibliography}
\end{document}